\documentclass{jfm}

\usepackage{color}
\usepackage{graphicx}
\usepackage{subfigure}
\usepackage{natbib}
\usepackage{epstopdf}
\usepackage{amsmath,amsfonts,amssymb}   %% AMS mathematics macros
\DeclareMathOperator*\Res{Res}

\definecolor{red}{rgb}{0,0,0}

\newcommand{\be}{\begin{equation}}
	\newcommand{\ee}{\end{equation}}
\newcommand{\bs}{\begin{subequations}}
	\newcommand{\es}{\end{subequations}}

\newcommand{\bi}[1]{{\boldsymbol #1}}
\newcommand{\bd}[1]{{\boldsymbol #1}}
\newcommand{\rmd}{\mathrm{d}}
\newcommand{\rme}{\mathrm{e}}
\newcommand{\rmi}{\mathrm{i}}

\newcommand{\bxi}{{\boldsymbol \xi}}

\newcommand{\half}{{\textstyle\frac12}}

\newcommand{\exom}{\rme^{-\rmi\Omega_0T}}
\newcommand{\exKR}{\rme^{\rmi\bK\cdot\bR}}
\newcommand{\Pext}{P_\mathrm{ext}}

\newcommand{\Fr}{\mathrm{Fr}}
\newcommand{\Frs}{\mathrm{Fr}_\mathrm{s}}

\newcommand{\Frsb}{\mathrm{Fr}_\mathrm{sb}}

\newcommand{\bx}{\bi{x}}
\newcommand{\bk}{\bi{k}}
\newcommand{\bK}{\bi{K}}
\newcommand{\bR}{\bi{R}}

\newcommand{\bV}{\bi{V}}

\newcommand{\bcdot}{\bi{\cdot}}

\newcommand{\kint}{\int\frac{\rmd^2 k}{(2\pi)^2}}

\newcommand{\gint}{\int^{\pi}_{-\pi} \rmd \gamma}

\newcommand{\efac}{\rme^{\rmi ( \bk \bcdot \bxi-\omega_0 t)}}

\newcommand{\hu}{\hat{u}}
\newcommand{\hv}{\hat{v}}
\newcommand{\hw}{\hat{w}}
\newcommand{\hp}{\hat{p}}

\newcommand{\pext}{p_\text{ext}}

\newcommand{\gR}{\gamma_ \text{excl}}
\newcommand{\tR}{\tau_\text{Res,min}}

\newcommand{\kV}{\bk \bcdot \bV}

\newcommand{\barr}{\left(\begin{array}{c}}
\newcommand{\earr}{\end{array}\right)}

\newcommand{\taures}{\tau_\mathrm{Res}}
\newcommand{\gammares}{\gamma_\mathrm{Res}}
\newcommand{\tgR}{\tilde{\gamma}_\mathrm{excl}}
\ifCUPmtlplainloaded \else
\checkfont{eurm10}
\iffontfound
\IfFileExists{upmath.sty}
{\typeout{^^JFound AMS Euler Roman fonts on the system,
		using the 'upmath' package.^^J}%
	\usepackage{upmath}}
{\typeout{^^JFound AMS Euler Roman fonts on the system, but you
		dont seem to have the}%
	\typeout{'upmath' package installed. JFM.cls can take advantage
		of these fonts,^^Jif you use 'upmath' package.^^J}%
	\providecommand\upi{\upi}%
}
\else
\providecommand\upi{\upi}%
\fi
\fi

\ifCUPmtlplainloaded \else
\checkfont{msam10}
\iffontfound
\IfFileExists{amssymb.sty}
{\typeout{^^JFound AMS Symbol fonts on the system, using the
		'amssymb' package.^^J}%
	\usepackage{amssymb}%
	  
	  \let\geq=\geqslant
}{}
\fi
\fi

\ifCUPmtlplainloaded \else
\IfFileExists{amsbsy.sty}
{\typeout{^^JFound the 'amsbsy' package on the system, using it.^^J}%
	\usepackage{amsbsy}}
{\providecommand\boldsymbol[1]{\mbox{\boldmath $##1$}}}
\fi

\title[Moving, oscillating surface disturbance on a shear current]{Multiple resonances of a moving, oscillating surface disturbance on a shear current}
\author[Y. Li and S. \AA. Ellingsen]{Yan Li$^1$\thanks{Email address for correspondence: yan.li@ntnu.no} and Simen \AA. Ellingsen$^1$}

\affiliation{$^1$Department of Energy and Process Engineering, Norwegian University of Science and Technology, N-7491 Trondheim, Norway}
\date{12 February 2016}           %% By default, LaTeX uses the current date

\begin{document}
\maketitle

\begin{abstract}
  We consider waves radiated by a disturbance of oscillating strength moving at constant velocity along the free surface of a shear flow which, when undisturbed, has uniform horizontal vorticity of magnitude $S$. When no current is present the problem is a classical one and much studied, and in deep water a resonance is known to occur when $\tau=|\bV|\omega_0/g$ equals the critical value $1/4$ ($\bV$: velocity of disturbance, $\omega_0$: oscillation frequency, $g$: gravitational acceleration). We show that the presence of the sub-surface shear current can change this picture radically. Not only does the resonant value of $\tau$ depend strongly on the angle between $\bV$ and the current's direction and the ``shear-Froude number'' $\Frs=|\bV|S/g$; when $\Frs>1/3$, multiple resonant values --- as many as $4$ --- can occur for some directions of motion. At sufficiently large values of $\Frs$, the smallest resonance frequency tends to zero, representing the phenomenon of critical velocity for ship waves. We provide a detailed analysis of the dispersion relation for the moving, oscillating disturbance, in both finite and infinite water depth, including for the latter case an overview of the different far-field waves which exist in different sectors of wave vector space under different conditions. Owing to the large number of parameters, a detailed discussion of the structure of resonances is provided for infinite depth only, where analytical results are available.
\end{abstract}
%%%%%%%%%%%%%%%%%%%%%%%%%%%%%%%%%%%%%%%%%%%%%%%%%%%%%%%%%%%%%%%%%%%%
%%%%%%%%%%%%%%%%%%%%%%%%% S E C T I O N %%%%%%%%%%%%%%%%%%%%%%%%%%%%
%%%%%%%%%%%%%%%%%%%%%%%%%%%%%%%%%%%%%%%%%%%%%%%%%%%%%%%%%%%%%%%%%%%%
\section{Introduction}

The problem of a wave source which is at one time oscillating and moving with respect to the free surface is a classical one, and a considerable literature exists when no shear current is assumed. 
{A key motivation for studying such a periodic travelling wave maker is its close mathematical relation to the classical problem of ship seakeeping in regular waves. 
The travelling oscillating source problem was first solved to linear order assuming purely oscillatory motion by \citet{haskind46} and extended by \citet{brard48}, \citet{eggers57}, \citet{havelock58} and others, when the perturbation is assumed to be from a submerged oscillating source; see also the review in \S13 of \citet{wehausen60}. When the water depth is finite, the analysis is richer, and was given by \citet{becker58}. The submerged source model is particularly useful since it doubles as a Green's function which may be used to describe the motion of floating vessels in waves \citep{newman59}. \citet{tayler74} used the ray method to study the linearised finite depth problem, allowing also a constant acceleration. The corresponding two-dimensional problem of a moving, oscillating line source was considered by \citet{haskind54} and extended to finite water depth by \citet{becker56}.

A much considered alternative model for a moving wave-maker is that of a pressure distribution at the fluid surface. A pressure distribution of static shape and uniform motion was used by \citet{havelock08} to study ship waves, and various cases of a pressure distribution in rectilinear motion which is simultaneously oscillating in strength were considered to linear order in two dimensions by \citet{wu57,kaplan57, debnath69}, and in three dimensions by \citet{lunde51,debnath69b}, and by \cite{doctors78} who applied it to wave resistance calculations for an air cushion vehicle. General considerations with a variety of applications were given by \cite{lighthill70}.

When the focus is on the dispersive properties of waves (rather than, say, wave-body interactions), the surface pressure model has the virtue of acting at the surface only, introducing no singular flow features in the interior of the liquid phase. When one assumes, as we do herein, that a shear current is present beneath the free surface, this benefit becomes particularly simplifying, because it was recently shown by \cite{ellingsen15a,ellingsen15b} that a submerged oscillating source in a rotational flow will generate a downstream series of vertical flow structures akin to a critical layer, with an accompanying ``critical wave''. We shall show herein that the introduction of a shear current increases the richness of the dispersion problem greatly, and eschewing additional complications from critical layer-like flow was deemed wise.

A particular feature of the oscillating and moving wave-maker is that a critical frequency-velocity combination exists at which resonance occurs, and wave amplitudes as predicted by inviscid linearised theory can become unbounded. For water gravity waves on deep, still waters the resonance is known to occur at $\tau=1/4$, where
\be\label{tau}
  \tau = \frac{\omega_0 V}{g},
\ee
$\omega_0$ is the angular frequency of the source, $V$ is its velocity relative to the surface, and $g$ is the gravitational acceleration. The resonance is explained by noting that the group velocity of waves emitted in the forward direction tends to zero relative to the moving source, hence wave energy is unable to propagate away. The resonant value is determined by the dispersion relation only, independently of the size, shape and nature %(source or pressure distribution) 
of the model wave maker, and is found \citep[e.g.,][]{tayler74} to be the same in 2D and 3D. It was shown by \cite{dagan80} that for the surface pressure source the wave amplitude diverges as $(\tau-1/4)^{-1/2}$ in 2D, and as $\ln(\tau-1/4)$ in 3D, for linearised, inviscid flow. The amplitudes become bounded once higher order terms are considered \citep{dagan80,dagan82}, being cancelled by 3rd order terms. For an extended submerged two-dimensional cylinder undergoing small oscillations, \cite{grue85} find radiated amplitudes to remain finite as $\tau\to 1/4$, even though the Green's function (describing waves from a point source) is known to diverge. \citet{liu93} showed that linearised wave amplitudes from extended bodies undergoing small oscillations in fact have a finite value at $\tau=1/4$ for all fully submerged bodies, whether 2D or 3D, as long as the body has a non-zero volume. 
Using a Rankine panel method, \cite{kring98} came to the same conclusion for a floating vessel. 

We show herein that when a sub-surface shear current is present, the resonant value of $\tau$ can change radically and in a non-trivial way. Indeed %, when a shear current is present, 
multiple resonant values of $\tau$ can occur, whose values depend on the direction of motion relative to the shear current, and on the non-dimensional ``shear-Froude number'' $\Frs=VS/g$, where $S$ is the constant vorticity of the undisturbed shear current.

Although of finite amplitude in realistic situations, the computations of \cite{kring98}, validated by experiments by \cite{maruo83}, show that wave loads on a vessel in regular waves can increase sharply near the Doppler resonance, potentially affecting seakeeping and wave resistance significantly. A similar behaviour was reported for a submerged ellipsoidal cylinder by \cite{grue86}. For this reason, knowledge of the structure of resonant frequencies on shear currents is of practical as well as theoretical interest. Several situations exist where the effect of shear is likely to have practical consequences.  For example, velocity profiles measured in the delta of Columbia river \citep{kilcher10} show strong shear in the top few meters of the water column, quite enough to affect resonance frequencies significantly; the data set was used for surface wave dispersion analysis by \cite{dong12}. Strongly sheared currents near the surface were also measured by \cite{haines94} in a barred surf zone. 

For example, the data from \cite{kilcher10}  show approximately linear shear of about $0.4$s$^{-1}$ for the top $3$m of the water column. For a small ship moving at $20$ knots, say, the resonant frequency of encounter from incoming waves can be roughly estimated to be increased by about a factor $2$ for motion in the most shear assisted direction ($\beta=\pi$ in figure \ref{fig:coordinate1}) and reduced by about a factor $1/3$ in shear inhibited direction ($\beta=0$). Even in much weaker shear currents the effect can be significant. 
A ship running at a preferred velocity so that $\tau>1/4$ might be surprised to suddenly hit resonant conditions when encountering such a shear current, but might avoid this by changing course, thereby changing the resonant frequency.

Other scenarios where shear will affect wave properties include shallow rivers, on which wavelengths long enough to be affected by the entire water column will notice the strongly sheared bottom boundary layer \citep[][\S IV]{peregrine76}. The results herein could equally well apply to wave loads on fixed or moored objects in rapid streams, where local shear can be very strong.

Of course, naturally occurring shear currents are not generally linear functions of depth as assumed herein. This assumption is made for simplicity. 
allowing one 
to analyse more clearly the effect of vorticity upon resonance frequencies, in our view a necessary step before embarking on more complicated realistic scenarios.

A phenomenon closely related with the Doppler resonance is waves generated by a ship near the critical velocity where its transverse waves vanish; it was shown that such a critical velocity, well known for shallow water waves, exists also in deep waters when a shear current is introduced \citep{ellingsen14a}, and in the presence of both shear current and finite water depth the critical situation depends on both factors \citep{li16}. Nonlinear wave phenomena become important at the critical velocity, where solitons may be produced \citep[e.g.,][]{ertekin86}. We show herein that this critical velocity occurs when, for $\Frs$ greater than a critical value, the smallest of the resonant values of $\tau$ drops to zero. 

We shall be concerned primarily with how the presence of a shear current modifies the resonant value(s) of $\tau$, and work to linear order in wave amplitude. We show that wave amplitudes diverge as $\ln(\tau-\taures)$ for $\tau$ approaching a resonant value $\taures$, in agreement with \cite{dagan80} except when $\taures=0$. The question of the finiteness or otherwise of wave amplitudes at resonance when a shear current is present, is a question for a later occasion.

%%%%%%%%%%%%%%%%%%%%%%%%%%%%%%%%%%%%%%%
%%%%%%%%%% S E C T I O N %%%%%%%%%%%%%%
%%%%%%%%%%%%%%%%%%%%%%%%%%%%%%%%%%%%%%%
\section{Formulation and general solutions}
We consider a three-dimensional wave-current system, incompressible and of negligible viscosity and surface tension. Our coordinate system is chosen so that surface velocity is zero, and the subsurface current is assumed to be aligned with the  $ x $ axis and vary linearly with depth according to the expression $ U(z)=Sz$ where $S$ is the uniform vorticity. Without loss of generality we assume $ S\geq0 $. The water has constant depth $h$. The free surface is disturbed by an applied external pressure distribution which moves with constant speed $V$ in a direction which makes an angle $\beta$ with the $x$ axis, and oscillates in strength around $0$ at a single frequency $\omega_0$. The disturbance from the pressure distribution is assumed to be sufficiently small that all equations of motion as well as boundary conditions may be solved to linear order in perturbation quantities. A sketch of the system is shown in Fig. \ref{fig:coordinate1}.

For our model, the velocity and pressure distribution are 
$\bd{v}= \left( U(z)+ \hu, \hv, \hw\right)$ and $P=\hp-\rho gz$ 
in which $ P $ is the total pressure, and $ \hp $ is the dynamic perturbation pressure; all hatted quantities are small perturbations due to existence of waves.
The solution of the linearised Euler equation proceeds in a similar fashion to that of \citet{ellingsen14b,ellingsen14a,li16} with the difference that we seek solutions which are purely oscillating in a coordinate system where the moving pressure source is at rest; for derivation details the reader may refer to these references.

%%%%%%%%%%%figure%%%%%%%%%%%%
\begin{figure}
	\graphicspath{{figures/}}
	\centering{\includegraphics[width=3in]{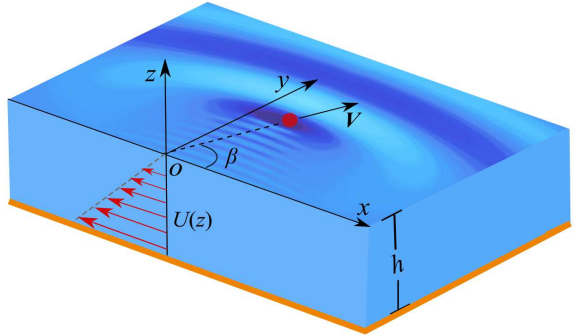}}	
	\caption{Geometry: a surface pressure distribution of oscillating strength travels with velocity $\bV$, making an angle $\beta$ with the $\bi{x}$ axis. A depth dependent shear current $U(z)=Sz$, parallel to the $\bi{x}$ axis, is present beneath the surface.}
	\label{fig:coordinate1}
\end{figure}
%%%%%%%%%%%%%%%%%%%%%%%%%%%%%

We may make the ansatz that all perturbation quantities will depend on time and space only through the Galilei transformed coordinate vector $ \bxi=\bx-\bV t $ (where $\bx=(x,y)$ is the horizontal position in the fixed coordinate system where the undisturbed surface is at rest) and an overall oscillating factor $\exp(-\rmi \omega_0t)$. We perform a Fourier transformation in the $\bxi$ plane according to
\be
[\hu,\hv,\hw,\hp](\bxi,z,t)  = 
\kint\efac[u,v,w,p](\bk,z) . 
\ee
in which $ \bk=(k_x,k_y)=(k\cos\theta,k\sin\theta) $ is the wave vector in the horizontal plane. We let $k=|\bk|$ be the wave number.

We will refer to source motion along directions $|\beta|<\pi/2$ in Fig.~\ref{fig:coordinate1} as ``shear inhibited" motion, being countered by the shear current compared to the case when no shear is present. Directions  $|\beta|>\pi/2$ we term ``shear assisted''. For a plane wave component of wave vector $\bk$ the same terminology is used for $|\theta|<\pi/2$ and $|\theta|>\pi/2$, respectively. This avoids the ambiguous terms ``upstream'' and ``downstream'' used the literature previously.

We then follow the same steps as in \citet{ellingsen14a} and \citet{li15a}, finding general solutions to the linearised continuity and Euler Equations under the bottom condition $w(\bk,-h)=0$, and compute undetermined coefficients by insertion into the kinematic and dynamic boundary conditions at the free surface in order to express the surface elevation in Fourier form. 

Let all lengths be scaled by the characteristic length of the moving source, $b$, and all times be scaled by $\sqrt{b/g}$. The surface elevation may be written in terms of non-dimensional quantities as
\be \label{eq:gxi}
\zeta (\bR,T)/b=\exom\gint\int_{0}^{\infty}\rmd K \frac{ \Pext(\bK) K^2 \tanh KH }{\Omega_+\Omega_-} \exKR. 
\ee
The dimensionless quantities which appear are, explicitly, 
\begin{align}
&\Omega_0=\omega_0\sqrt{b/g} , ~~   \bK=b\bk , ~~ T=t \sqrt{g/b}, ~~  H=h/b , ~ ~ \Fr=V/\sqrt{gb}; \notag \\
&\Frs=VS/g,~~ \bR = \bxi/b, ~~ \Pext(\bK) = \pext(\bk)/(\rho g b^3),
\end{align}
where $\Fr$ is the Froude number based on a length $ b $ 
and $ \Frs$ is the ``shear-Froude number" based on  the ``shear length" $ g/S^2 $. Definitions of the different angles involved, including $ \theta $, $ \beta$, $ \gamma $ and $\phi$, are shown in Fig.\ref{fig:angels1}. 
$\pext(\bk)$ is the Fourier transformed external pressure distribution. 
The non-dimensional physical quantities $ \Omega_\pm $ are defined as 
\bs
\begin{align}
&\Omega_\pm(\bK)=\omega_\pm \sqrt{\dfrac{b}{g}}=\Omega_0+K\Fr\cos \gamma-\Sigma_\pm(K,\gamma), \label{eq:def_omega_pm}\\
& \Sigma_\pm (\bK)=\sigma_\pm\sqrt{\dfrac{b}{g}}= \pm \sqrt{K\tanh KH + (\half \Frsb\cos \theta \tanh KH)^2}-\half \Frsb\cos \theta \tanh KH\label{eq:def_intri_freq}
\end{align}
\es
where we have introduced the ``intrinsic shear-Froude number'' 
\be
\Frsb=S\sqrt{b/g} = \Frs/\Fr. 
\ee
$\Sigma_\pm (\bK)$ are the non-dimensional intrinsic frequencies for a wave vector $\bK$.

%%%%%%%%%%%figure%%%%%%%%%%%%
\begin{figure}
	\graphicspath{{figures/}}
	\centering{\includegraphics[width=0.45\textwidth]{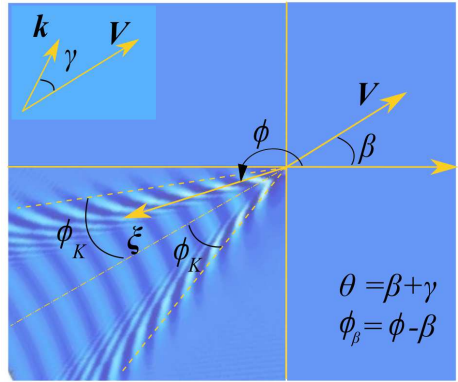}}
	\caption{Definition of angles. $\theta$ is the angle of the wave vector in the $xy$ coordinate system, $\gamma$ is relative to the sources direction of motion. $\phi_\beta$ is the angle between $\bV$ and spatial position $\bxi$ in the coordinate system where the disturbance is at rest.}
	\label{fig:angels1}
\end{figure}
%%%%%%%%%%%figure%%%%%%%%%%%%

We have neglected viscous damping on the basis that this plays a very minor role for linear gravity waves
in the absence of shear. Quite to what extent this holds also in the presence of vorticity, where dissipation could be stronger, is an open question. In the present context viscosity would prevent the occurrence of infinite wave amplitudes in a linearised theory, although for water waves at resonance viscous damping would likely be a weak effect compared to nonlinear corrections, which render wave amplitudes finite in any case.

%%%%%%%%%%%%%%%%%%%%%%%%%%%%%%%%%%%%%%%%%%%%%%%%%%%%%%%%%%%
%%%%%%%%%%%%%%%%%%%%%%%%% S E C T I O N %%%%%%%%%%%%%%%%%%%%%%%%%%%%
%%%%%%%%%%%%%%%%%%%%%%%%%%%%%%%%%%%%%%%%%%%%%%%%%%%%%%%%%%%%%%%%%%%%
\section{Dispersion relation analysis}

In order to be a nontrivial solution of the linearised Euler equations, plane wave components of wave vector $\bk$ must satisfy an eigenvalue condition posed by the boundary conditions, the dispersion relation. When the oscillation frequency is fixed to a value $\omega_0$ as here, there may be zero, one or two solutions (``waves'') of different $k$ for each phase propagation angle $\theta$. The wave field far from the oscillating pressure source is well known to consist only of the waves satisfying the dispersion relation; see e.g. \S4.9 of \cite{lighthill78} or \cite{lighthill70}. 

The dispersion relation for the 3D system in the presence of a shear flow of uniform vorticity might first have been derived by \citet{charland12}. It follows from 
the free surface boundary conditions 
and coincides with the position of poles of the integrand in Eq.~\eqref{eq:gxi}, that is, $\Omega_+\Omega_-=0$, which may be written
\begin{equation}
\omega_0 + \kV = \pm \sqrt{gk \tanh kh + (\half S\cos \theta \tanh kh)^2}-\half S\cos \theta \tanh kh=\sigma_\pm(\bk). 
\label{eq:disp-relation}
\end{equation}
In our discussion we will mainly use the nondimensional form of the dispersion relation, 
\be  \label{eq:disp_r}
\Omega_0+K\Fr\cos \gamma=\Sigma_\pm(\bK).
\ee

%%%%%%%%%%%%%%%%%%%%%%%%%%%%%%%%%%%%%%%%%%%%%%%%%%%%%%%%%%%%%%%%%%%%%%%%%%%%%%%
\subsection{Intrinsic group and phase velocity}
In the following we will discuss in detail solutions of the dispersion relation in different directions of plane wave propagation.
The discussion follows the same lines as that in \S3.7.1 of \citet{mei05}.

The intrinsic group velocity of a plane wave of wave vector $\bk$ is given by the vectorial quantity
\be
\bi{c}_g(\bk) = \nabla_k\sigma_\pm(\bk)
\ee
where $\nabla_k=(\partial/\partial k_x,\partial/\partial k_y)$ is the gradient operator in the $\bk$ plane, and $\sigma_\pm$ is defined in \eqref{eq:def_intri_freq}. In particular, $\partial \sigma_\pm(\bk)/\partial k$ is the component of $\mathbf{c}_g$ along direction $\bk$, which we term the radial component. 
The intrinsic frequencies and velocities are independent of the motion of the source and are velocities measured in the ``lab'' frame of reference, i.e., where the fluid surface is at rest. Relative velocities are measured relative to the moving source. The relative group velocity is 
\be
  \bi{c}_g^\mathrm{R}(\bk)=\bi{c}_g(\bk) - \bV=\nabla_k(\sigma_\pm-\bk\cdot\bV). 
\ee

Likewise, the intrinsic phase velocities are 
$\sigma_\pm(\bk)/k$, 
one of which positive, the other negative. A negative intrinsic %group or 
phase velocity for a wave component $\bk$ means that wave 
phase 
is
transported in direction $-\bk$. On the other hand the relative phase velocity of a wave $\bk$ is 
$\sigma_\pm(\bk)/k-V\cos\gamma$ 
which, by virtue of the dispersion relation \eqref{eq:disp-relation}, equals $\omega_0/k$ and is necessarily positive, hence the relative phase velocity is always directed along $\bk$. The relative group velocity, however, can have a negative component along direction $\bk$. 

In non-dimensional terms, the intrinsic 
and relative 
group velocities are
\be\label{cg}
  \bd{C}_g(K,\gamma) = \nabla_K\Sigma_\pm(\bK); ~~ ~~ \bd{C}_g^\mathrm{R}(K,\gamma) = \nabla_K(\Sigma_\pm(\bK)-K\Fr\cos\gamma).
\ee
Especially the radial component, $\partial_K \Sigma_\pm$, will be important in the following discussions.

%%%%%%%%%%%%%%%%%%%%%%%%%%%%%%%%%%%%%%%%%%%%%%%%%%%%%%%%%%%%%%%%%%%%%%%%%%%%%%%
\subsection{Dispersion relation in finite water depth}

In the following we consider graphical solutions of the dispersion relation $\Omega_\pm(\bK)=0$. In our theory, physical quantities are given by Fourier integrals over the whole $\bk$ plane, such as Eq.~\eqref{eq:gxi}, and far from the source only the waves satisfying the dispersion relation will be present \citep[cf., e.g.,][]{lighthill78}. A shear current introduces a richness of solutions which we find it necessary to discuss carefully and in some detail. 

We shall first consider the case of finite depth which, despite being formally a little more cumbersome, is more straightforward in principle than the deep water case. 

In each case we proceed by solving the dispersion relation \eqref{eq:disp_r} graphically by considering intersections between the straight lines $\Omega_0+\Fr\cos\gamma$ and the curves $\Sigma_\pm(K,\gamma)$ in different circumstances. Each intersection corresponds to a possible far-field wave, and the graphs in Figs.~\ref{fig:dispReV0} to \ref{fig:dispRe-inf} afford some immediate physical insights: Considering an intersection $P$ (say) for one particular propagation direction $\gamma$, the (non-dimensional) intrinsic phase velocity of the corresponding far-field wave is the slope of the 
straight 
line connecting P to the origin, the radial component of the intrinsic group velocity is the slope of the tangent of the curve $\Sigma_\pm$ at point $P$, and the velocity of source motion projected onto direction $\bk$ is the slope of the 
straight 
line connecting P to the point ($0,\Omega_0$). 

%%%%%%%%%figure%%%%%%%%%
\begin{figure}
	\graphicspath{{figures/}}
	\centering{\includegraphics[width=.75\textwidth]{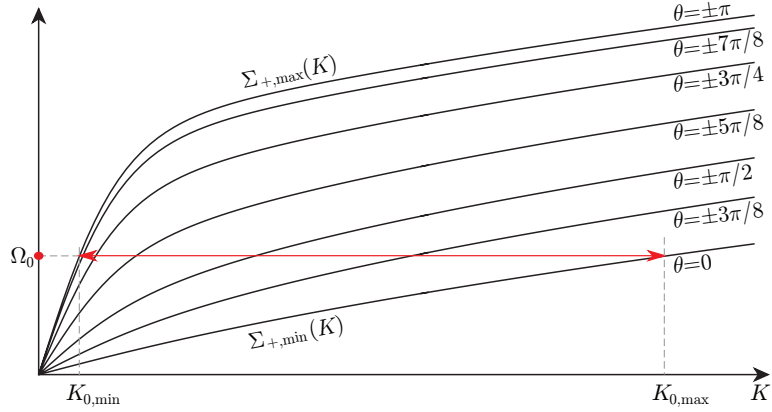}}
	\caption{Graphical solutions of Eq.~\eqref{eq:disp_r} when $V=0$.}
	\label{fig:dispReV0}
\end{figure}
%%%%%%%%%figure%%%%%%%%%

\subsubsection{Stationary source, $V=0$}

We begin by regarding the dispersion relation \eqref{eq:disp_r} when the wave source is at rest with respect to the surface. Although the nature of the wave source is different, the dispersion properties will be identical to the propagating waves from the stationary submerged oscillating source considered by \citet{ellingsen15b}. 

When $\Fr=0$, Eq.~\eqref{eq:disp_r} reduces to $\Omega_0=\Sigma_+(K,\theta)$. Graphical solutions are sketched in Fig.~\ref{fig:dispReV0} where $\Sigma_+(K,\theta)$ is plotted as a function of $K$ for different angles $\theta$. In the finite water case there is always a single far-field wave in all directions. For angles $\theta=\pm\pi/2$ the intrinsic phase velocity is 
unaffected by the shear. 
The maximal solution, $K=K_{0,\mathrm{max}}$, is for $\theta=0$ (wavelength shortened compared to no shear) and the minimum $K=K_{0,\mathrm{min}}$ (wavelength elongated by shear) is found at $\theta=\pm\pi$. Since the curve $\Sigma_+$ is everywhere concave down, intrinsic group velocity is always smaller than intrinsic phase velocity.

%HERE HERE

\subsubsection{Moving source, $V>0$}

A far richer situation ensues once the source is in motion relative to the water surface, $\Fr>0$. The situation is sketched in Fig.~\ref{fig:dispRe}a. Which far-field waves now exist for different propagation angles can vary strongly as a function of propagation direction $\gamma$. The figure is to be understood in a qualitative sense since the curves $\Sigma_\pm(K,\theta)$ also depend on $\gamma$ when the direction of motion, $\beta$, is fixed, but since the sign of the curvature of these graphs remains the same for all $\theta$ (see Fig.~\ref{fig:dispReV0}), the sketch is sufficient to visualise in a qualitative way the possible cases that occur. 

%%%%%%%%%figure%%%%%%%%%
\begin{figure}
	\graphicspath{{figures/}}
	\centering{\includegraphics[width=\textwidth]{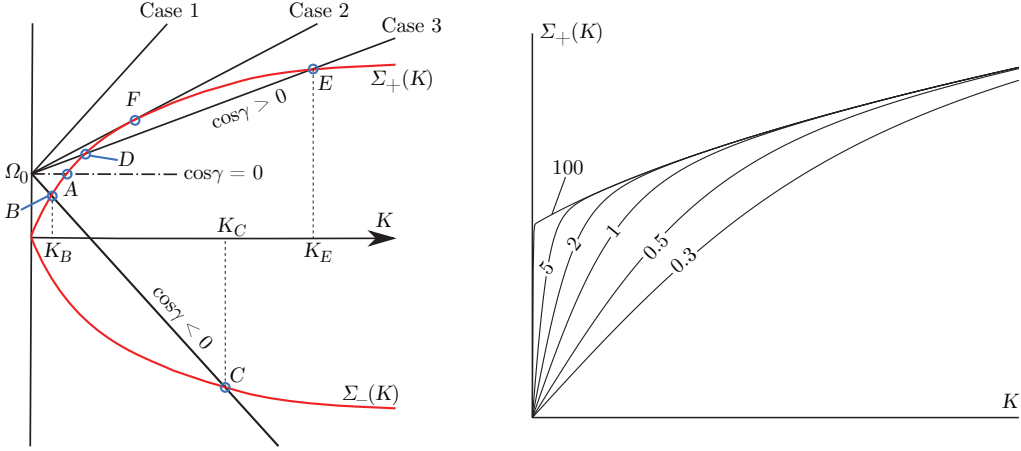}}
	\caption{a: Graphical solutions of Eq.~\eqref{eq:disp_r} in different situations. See text for discussion and details. b: Behaviour of $\Sigma_+(K)$ as $H$ tends to infinity, when $\cos\theta<0$ is assumed ($\Frsb\cos\theta=-2$ in the figure). Values of $H$ shown as numbers on respective graphs.}
	\label{fig:dispRe}
\end{figure}
%%%%%%%%%figure%%%%%%%%%

\paragraph{Situation 1: Orthogonal waves ($\cos\gamma=0$).}
The source velocity projected onto $\bk$ is now zero, and the situation is as for $V=0$. There is always a single solution to the dispersion relation, point A in Fig.~\ref{fig:dispRe}a. 

\paragraph{Situation 2: Sternward waves ($\cos\gamma<0$).} 
Since $\bk\bcdot\bV<0$, we denote this situation somewhat roughly as having ``rearward pointing'' wave vectors. There are two solutions corresponding to the points $ B $ and $ C $ in Fig. \ref{fig:dispRe}a. When comparing to situation 1 ($ \Fr\cos\gamma =0 $ at point $ A $), the wave represented by $ B $ is lengthened since $ K_B<K_A $, and intrinsic phase and group velocities are increased,
and a second solution $C$ also occurs with short wavelength and smaller (absolute) velocity.
The slope of the straight line intersecting $B$ and $C$ is negative, indicating that source velocity $\bV$ has a negative component along direction $\bk$. Wave $B$ thus travels rapidly in rearward directions. The wave corresponding to point $C$ has negative intrinsic phase and group velocities meaning that although it is emitted in a rearward direction by the source, it is seen in the lab system to propagate in a forward direction. A $C$-wave of wave vector $\bk$ has intrinsic phase velocity along direction $-\bk$. 

\paragraph{Situation 3: Forward waves ($ \cos\gamma>0 $).} 
This situation corresponds to ``forward directed'' wave vectors, and is the most complicated situation. The pertinent solutions are  $ \Sigma_{0+}=\Omega_0 +\Fr K\cos\gamma$, and there are now three different sub-cases as illustrated in Fig.\ref{fig:dispRe}a. When $\Fr\cos\gamma$ is sufficiently large, no waves exist (Case 1 in Fig.~\ref{fig:dispRe}). This situation can occur provided the non-dimensional frequency parameter 
$\tau = \Omega_0\Fr$, defined in Eq.~\eqref{tau}, 
exceeds a critical value $\tR$, the smallest Doppler resonant frequency, to be discussed 
in section \ref{sec:taumin}. 
At a critical value of $\Fr\cos\gamma$ only a single wave $F$ of wave number $K_F$ exists (Case 2), corresponding to a double root. The critical values $\gamma=\gR$ where this occurs can be found by noting that radial group velocity equals projected source velocity at this point:
\be \label{eq:criticalSituation}
\frac{\partial\Sigma_+}{\partial K}(K_F,\gR)=\Fr\cos\gR
\ee
where the notation means the derivative is evaluated at point $(K_F,\gR)$.
For supercritical values of $\tau$ there exists at least one sector $\gR^-<\gamma < \gR^+$ where Case 1 occurs, although as we shall detail in the case of deep water, as much as three such exclusion sectors may exist, depending on $\beta$ and $\Frs$. 

For $ \cos\gamma $ smaller than $\cos\gamma_\text{excl}$ there are two possible solutions corresponding to points $ D $ and $ E $, denoted Case 3 in Fig.~\ref{fig:dispRe}. When $\tau<\tR$ this is the only possible case for $\cos\gamma>0$. Point $D$ corresponds to the faster and longer of the two waves, and since its radial group velocity is greater than the source velocity projected onto the same direction, $\bk$, this wave is the only one which might be found in front of the moving source. Wave $E$ has shorter wavelength and moves more slowly, and though propagating in a forward direction, is left behind by the moving source.

In Fig.~\ref{fig:dispRe}b we illustrate the behaviour of the intrinsic frequency $\Sigma_+$ as the water depth $H$ increases; in the figure $\Sigma_+(K,\theta)$ is plotted as a function of $K$, presuming $\cos\theta < 0$ (for $\cos\theta>0$ the situation is mirrored about the abscissa). As $H\to\infty$, the graph of $\Sigma_+(K)$ obtains the shape of a straight vertical line from the origin to the point $-\Frsb\cos\theta$ ,
thence following a curved shape concave towards the abscissa. This behaviour, leading to the phenomenon of cutoff as discussed in \citet{tyvand15,ellingsen15a,ellingsen15b}, becomes important when next considering the deep water case.

%%%%%%%%%%%%%%%%%%%%%%%%%%%%%%%%%%%%%%%%%%%%%%%%%%%%%%%%%%%%%%%%%%%%%%%%%%%%%%%
\subsection{Dispersion relation in infinite water depth}
When assuming water depth to be infinite, the situation becomes at one time both simpler, in that explicit solutions to the dispersion relation may now be found, and more complicated. Concerning the graphical solutions of the dispersion relation, Eq.~\eqref{eq:disp_r}, the curve of $\Sigma_+$ or $\Sigma_-$ obtains a vertical section from the origin to value $-\Frsb\cos\theta$ as illustrated in Fig.~\ref{fig:dispRe}b. The graphical solution situation for infinite water depth is shown schematically in Fig.~\ref{fig:dispRe-inf}. We will distinguish between what we term \emph{weak} and \emph{strong shear} situations. 

%%%%%%%%%%%figure%%%%%%%%%%%%
\begin{figure}
	\graphicspath{{figures/}}
	\centering
	\subfigure{\includegraphics[width=\textwidth]{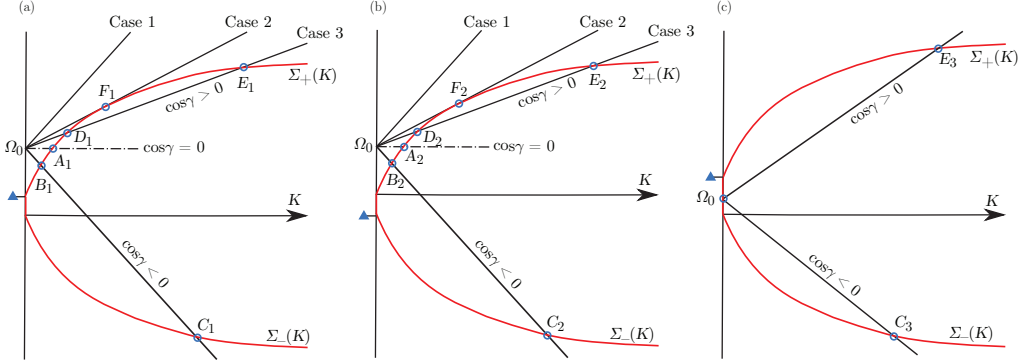}}
	\caption{Graphical solutions of dispersion relation \eqref{eq:disp_r} in infinite water depth. Intrinsic frequencies $\Sigma_\pm(K,\gamma)$ are the curved lines, $K\Fr \cos\gamma$ are the straight lines. Intersections of the two are wave solutions, marked with circles. The point $(0,-\Frsb\cos\theta)$ is marked with a triangle. The panels show situations $0<-\Frsb\cos\theta<\Omega_0$ (a), $-\Frsb\cos\theta<0$ (b), $0<\Omega_0<-\Frsb\cos\theta$ (c). }
	\label{fig:dispRe-inf}
\end{figure}
%%%%%%%%%%%figure%%%%%%%%%%%%

\paragraph{Weak shear.}
When  $\Omega_0 > \Frsb$, or equivalently, $\omega_0>S$, the discussion of which far-field waves occur is qualitatively identical to that for finite water depth given above. %The situation can occur with either weak shear or high frequency. 
Only the situations shown in Fig~\ref{fig:dispRe-inf}a and \ref{fig:dispRe-inf}b can occur in this case.

\paragraph{Strong shear.}
When $\omega_0<S$, however,  the appearance of the vertical section of the graph of $\Sigma_+$ (along which $K=0$) when $\cos\theta<0$ means that a new situation will arise, not found in finite water depth. In this, strong shear case, a sector of angles $\theta$ exists centred at $\theta=\pi$, within which waves of type $A,B$ and $D$ all have wave number $K\to 0$ as $H\to\infty$. The situation  within this sector is depicted in Fig~\ref{fig:dispRe-inf}c.  This is the phenomenon of ``cut-off'' discussed for the 2D case in \citet{tyvand14,tyvand15,ellingsen15a} and briefly in 3D with $V=0$ in \citet{ellingsen15b}. It is shown in \citet{ellingsen15a} that these $K=0$ modes carry no energy and may simply be disregarded in the far-field, and we will not consider them to be solutions. Cut-off of $A,B$ and $D$ waves occurs in a sector $\pi-\theta_0<\theta<\pi+\theta_0$, where
\be
\theta_0 = \arccos(\omega_0/S).
\ee

The dispersion relation at infinite water depth may be found by taking $ KH \to \infty $ in \eqref{eq:disp_r},
\be \label{eq:dispR_inf}
\Omega_0+K\Fr\cos\gamma=\pm\sqrt{K + (\half\Frsb\cos \theta)^2}-\half\Frsb\cos \theta 
\ee
which is valid for $K$ strictly greater than $0$. Unlike for finite depth, this dispersion relation permits the analytical solutions
\bs \label{Kdeep}
\begin{align}
K_{C,E}=&\dfrac{1-\Frs\cos\theta\cos\gamma-2\tau\cos\gamma+\sqrt{\Delta}}{2\Fr^2\cos^2\gamma},\\
K_{B,D}=&\dfrac{1-\Frs\cos\theta\cos\gamma-2\tau\cos\gamma-\sqrt{\Delta}}{2\Fr^2\cos^2\gamma},
\end{align}
\es
with discriminant
\be \label{eq:Delta}
\Delta=(1-\Frs\cos\gamma\cos\theta)^2-4\tau\cos\gamma, 
\ee
which agree with \citet{wehausen60,dagan82} without shear current. Subscripts $ B,C,D $ and $ E $ refer to labels on graphical solutions in Fig. \ref{fig:dispRe-inf}; waves $B,C$ are obtained if $\cos\gamma<0$, waves $D,E$ if $\cos\gamma>0$. We shall make use of these solutions towards analysing Doppler resonances in section \ref{sec_doppler}.

%%%%%%%%%%%figure%%%%%%%%%%%%
\begin{figure}
	\graphicspath{{figures/}}
	\centering
	\includegraphics[width=\textwidth]{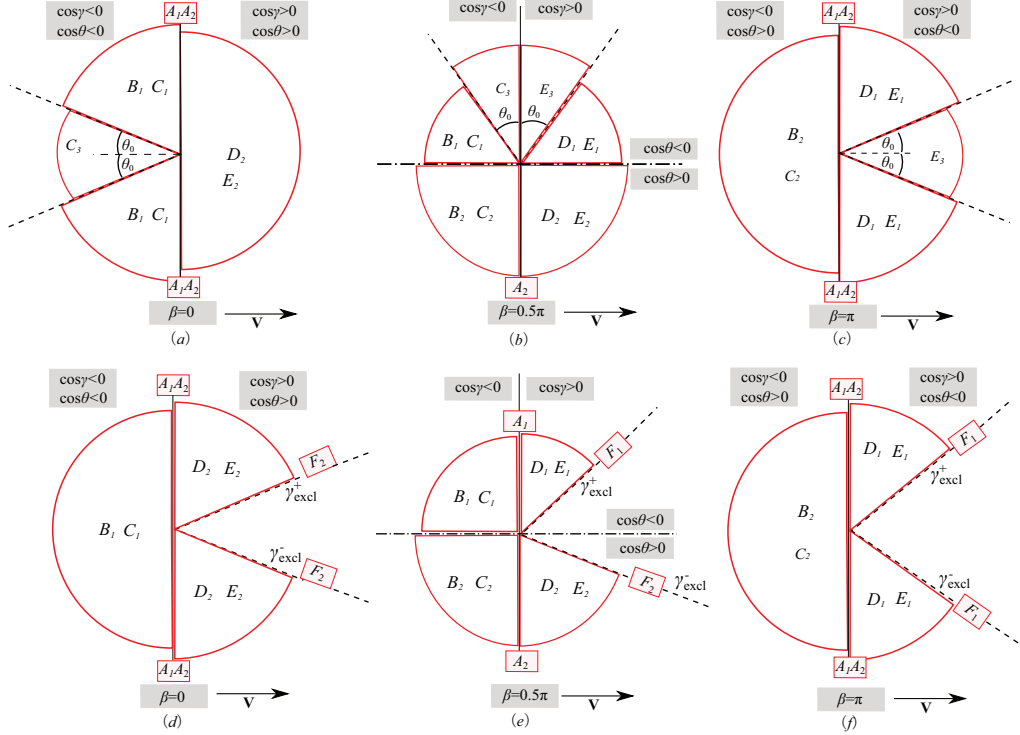} 
	\caption{Examples showing different far-field waves occurring in different sectors of the $\gamma$ plane for $\beta=0$ (a,d), $\beta=\pi/2$ (b,e) and $\beta=\pi$ (c,f). Above (a-c): $\tau<\tR$ and $\omega_0<S$. A cutoff sector of total angle $2\theta_0$ appears within which there are no $A,B$ or $D$ waves. Below (d-f): $\tau>\tR$ and $\omega_0>S$. For $\tau$ above the Doppler resonance, no waves exist within a sector of forward directions. In all figures the source's direction of motion, $\gamma=0$, is towards the right. Labels within each sector refer to graphical solutions of the dispersion relation as shown in Fig.~\ref{fig:dispRe-inf}}
	\label{fig:pie}
\end{figure}
%%%%%%%%%%%figure%%%%%%%%%%%%

In summary, which and how many far-field waves are found when propagation angle $\gamma$ (or $\theta$) is varied between $-\pi$ and $\pi$ is now determined by two criteria: Whether or not $\tau$ exceeds the smallest Doppler resonance frequency $\tR$, and whether or not $\omega_0$ exceeds $S$. With reference to Fig.~\ref{fig:dispRe-inf} we show some examples in Fig.~\ref{fig:pie} of which far-field waves appear in which sectors of the $\bk$ plane when $\gamma$ is varied through a full circle. In Fig.~\ref{fig:pie}a-c we assume $\omega_0<S$ and $\tau<\tR$, so there is a cut-off sector symmetrical about $\theta=\pi$. In Fig.~\ref{fig:pie}d-f we assume $\omega_0>S$ and $\tau>\tR$. Now there is no cut-off, but a sector $\gR^-<\gamma<\gR^+$ appears within which no propagating wave solutions exist since $\Delta<0$ in Eq.~\eqref{Kdeep}.

%%%%%%%%%%%%%%%%%%%%%%%%%%%%%%%%%%%%%%%%%%%%%%%%%%%%%%%%%%%%%%%%%%%%%%%%%%%%%%%
%%%%%%%%%%%%%%%%%%%%%%%%%%% S E C T I O N %%%%%%%%%%%%%%%%%%%%%%%%%%%%%%%%%%%%%
%%%%%%%%%%%%%%%%%%%%%%%%%%%%%%%%%%%%%%%%%%%%%%%%%%%%%%%%%%%%%%%%%%%%%%%%%%%%%%%
\section{Doppler resonance in deep water} \label{sec_doppler}

\subsection{Excluded sectors}

Before studying the phenomenon of Doppler resonance, we will regard sectors of propagation directions $\gamma$ in which $D$ and $E$ waves become evanescent and do not appear in the far-field. In deep water it is clear to see from the explicit solutions \eqref{Kdeep} that this situation is characterised by $\Delta<0$, that is,
\be\label{tauex}
\tau > \frac{[1-\Frs\cos(\gamma+\beta)\cos\gamma]^2}{4\cos\gamma}\equiv \Phi(\gamma).
\ee

\begin{figure}
	\graphicspath{{figures/}}
	\includegraphics[width = \textwidth]{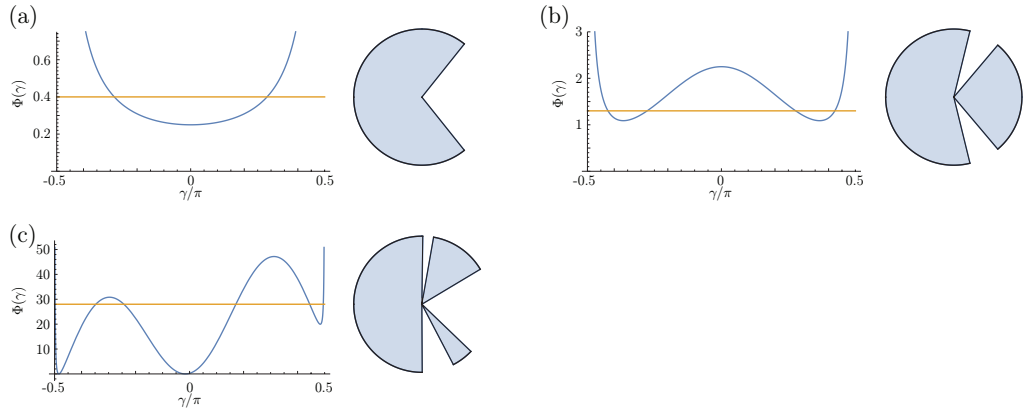}
	\caption{Sectors of propagation directions $\gamma$ with no wave solutions. The graphs are plots of $\Phi(\gamma)$, the horizontal lines show the (arbitrarily chosen) value of $\tau$; sectors where $\Phi(\gamma)<\tau$ are excluded. Pie charts next to each graph illustrate excluded sectors (included sectors are shaded). We show in Section \ref{sec:respos} that a resonant value of $\tau$ is associated with each local maximum or minimum of $\Phi(\gamma)$; if $\Phi(\gamma)$ has a local extremum at $\gammares$, then $\taures=\Phi(\gammares)$. (a) The case without shear, $\Frs=0$, $\tau=0.4$. (b) Moderate case with two excluded sectors, $\Frs=2, \beta=\pi$, $\tau=1.3$. (c) Extreme case with three excluded sectors, $\Frs=20, \beta=\pi/2$, $\tau=28$.}
	\label{fig:excl}
\end{figure}

The phenomenon is well known also without shear, in which case exclusion of a single sector $|\gamma|<\arccos(1/4\tau)$ occurs whenever $\tau>1/4$. In the presence of shear, however, there can be as much as three separate excluded sectors. Fig.~\ref{fig:excl}a shows the no-shear case with a single excluded sector. %However, even for 
For moderate values of $\Frs$, two excluded regions can occur, as illustrated in Fig.~\ref{fig:excl}b. 
Note that the excluded sectors do not include $\gamma=0$, so although $\tau>\tR$, the source still has a wave travelling ahead of it.
For large values of $\Frs$ it is even possible for a third sector to appear, as shown in Fig.~\ref{fig:excl}c. The values of $\Frs$ and $\tau$ involved are so large that this case is of doubtful practical significance.

%%%%%%%%%%%%%%% S E C T I O N %%%%%%%%%%%%%%%%%%
\subsection{Criterion of the Doppler Resonance}\label{sec:criterion}

Physically, a Doppler resonance refers to the case where wave energy is held stationary in space from the perspective of the moving source, i.e.\ zero relative group velocity. We obtain $\bd{C}_g^\mathrm{R}$
in a polar 
$\bd{k}$-coordinate system 
from Eq.~\eqref{cg} 
as 
\begin{align}
  \bd{C}_g^\mathrm{R} =& \left[\pm \half (K+\Fr^2_{sb}\cos^2\theta/4)^{-1/2}-\Fr\cos\gamma\right]\bd{e}_k \notag \\
  &+ \left[\Fr\sin\gamma+\dfrac{\Fr_{sb}\sin\theta}{2K}\mp\frac1{8K}\Fr^2_{sb}(K+\Fr^2_{sb}\cos^2\theta/4)^{-1/2}\sin2\theta \right]\bd{e}_\theta
\end{align}
The Doppler resonance occurs when $ \mathbf{C}^{\text{R}}_g=0 $ which, after eliminating $K$ may be written \be \label{psi}
\Phi'(\gamma)=\dfrac{(1-\Frs\cos\gamma\cos\theta)[(1-\Frs\cos\gamma\cos\theta)\tan\gamma + 2\Frs \sin(\gamma+\theta)]}{4\cos^2\gamma}=0
\ee
where $-\pi/2<\gamma<\pi/2$ is assumed. For each $\gamma$ solving Eq.~\eqref{psi} there exists a resonant (not necessarily distinct) value of $\taures$. We will see that this criterion is identical to the condition for infinite amplitudes to be possible, namely that $\Delta(\gamma)=\Delta'(\gamma)=0$.

%%%%%%%%%%%%%%% S E C T I O N %%%%%%%%%%%%%%%%%%
\subsection{Wave amplitude of $F$-type wave}

We will now consider the wave amplitude due to waves of type $D$ or $E$ (see Fig.~\ref{fig:dispRe-inf}) when these flow together in a single point $F$ at some value $\gamma=\gR$. 

The surface elevation \eqref{eq:gxi} is in the form of an integral over $\bk$. As is typical of wave descriptions with periodic or stationary time dependence, this integral is not well defined until a radiation condition has been applied. Using the procedure of, e.g., \S4.9 of \citet{lighthill78}, we replace
$\Omega_0\to\Omega_0+\rmi \epsilon$
where $\epsilon=0^+$, i.e., $\Omega_0$ is given a small positive imaginary part, so that 
\be\label{eq?radiC}
e^{-\rmi\Omega_0 T} \longrightarrow e^{-\rmi\Omega_0T+\epsilon T}.
\ee
The procedure is closely related to that for ship waves in \citet{li16}, where more detailed discussion may be found.
The introduction of the small quantity $\epsilon$ now moves the poles where $\Omega_\pm=0$ slightly into the complex $K$ plane, and the integral is well defined. 

Henceforth, let us assume infinite water depth to allow more explicit analysis. Using partial fractions 
\eqref{eq:gxi} may be written in the form
\bs \label{eq:fullf_1}
\begin{align}
\zeta(\bR,T)/b=&\exom \gint [I_+(\gamma)-I_-(\gamma)],\\
I_\pm(\gamma)=&\lim_{\epsilon\to 0}\int\limits_{0}^{\infty}\rmd K \frac{f(\bK)\rme^{\rmi \bK\cdot \bR}}{\Omega_\pm(K,\gamma)+\rmi\epsilon}, \\
f(\bK)=&\frac{\Pext(\bK)K^2}{2\sqrt{K+(\half\Frsb\cos \theta )^2}}.\label{f}
\end{align}
\es
The external pressure distribution is not specified, but we assume it is well localised so that the integral over $K$ converges. 

It is well established (e.g., \citet{lighthill78,li16}) that the leading order contribution to $\zeta$ far from the source comes from the contribution to the $\bK$ integral from the poles where $\Omega_\pm(K,\gamma)=0$. We call this contribution the far-field, and it consists of the waves of type $A$ to $E$ as shown in Fig.~\ref{fig:dispRe-inf} for deep water. 

We assume $\tau>\tR$, and consider $\gamma$ close to (but just outside) a sector wherein $\Delta<0$, as illustrated in Fig.~\ref{fig:excl}. 
For simplicity we assume that there is a single such sector delimited by $\gR^\pm$. In this sector, waves of type $D$ and $E$ become evanescent and do not contribute to the far-field. 
Since $\gamma$ is close to some $\gR$, far-field waves of type $D$ and $E$ in Fig.~\ref{fig:dispRe-inf} have almost, but not exactly, the same wave number, and the quantity $\Delta(\gamma)$ from Eq.~\eqref{eq:Delta} is small, but positive.
We want the leading order contributions to the far-field integrals as $\gamma$ approaches $\gR$ from the side where $D,E$ waves exist. $D$ and $E$-waves are forward-propagating ($\cos\gamma>0$) so far-field waves are solutions of $\Omega_+(K,\gamma)=0$. Since wave $D$ has group velocity greater than $V\cos\gamma$, this wave is found in the far-field in front of the source, $\bK\cdot\bR>0$, while the $E$ wave is found behind, so for a single position $\bR$ in the far-field only one of these waves can ever contribute, and the two waves do not interfere with each other [mathematically, the poles corresponding to $D$ and $E$ waves lie on opposite sides of the real $K$ axis]. 

We consider a $D$-wave (wave number $K_D$) for definiteness.
The far field $D$-wave surface elevation is now found as
\be\label{zetaD}
\zeta_{\mathrm{f.f.}}^D/b =\exom \left[\int_{\gamma_0}^{\gR^-}+\int_{\gR^+}^{\gamma_1} \right] \rmd \gamma I_+^D(\gamma)
\ee
where $I_+^D$ is now approximated by the contribution from the pole near $K=K_D$ only. The limits $\gamma_0$ and $\gamma_1$ are non-singular and
can give a finite contribution only.

Assume that the pole is simple so that 
\be
\Res_{K=K_D}\frac{f(\bK)\rme^{\rmi \bK\cdot \bR}}{\Omega_+(K,\gamma)} = \frac{f[K_D(\gamma),\gamma]\rme^{\rmi K_D(\gamma)R\cos(\gamma-\phi_\beta)}}{\Omega'_+[K_D(\gamma),\gamma]}
\ee
where $\phi_\beta$ is the angle between $\bR$ and $\bV$ and a prime denotes $\partial/\partial K$.
Define $K_F(\gamma)$ as the value of $K$ so that $\Omega'(K_F,\gamma)=0$, found by assuming $\Delta=0$ in \eqref{Kdeep},
\be
K_F(\gamma) = \frac{1-(\Frs \cos\gamma\cos\theta)^2}{4\Fr^2\cos^2\gamma}
\ee
which also solves $\Omega_+=0$ if $\gamma=\gR$. Then $  \Omega'_+(K_D,\gamma) \approx (K_D-K_F)\Omega''_+(K_F,\gamma)$, and evaluating the contribution from the residue of the pole at $K=K_D$ we obtain
\be
I_+^D(\gamma)\approx 2\pi\rmi \frac{f(K_F)e^{K_FR\cos(\gamma-\phi_\beta)}}{(K_D-K_F)\Omega''_+(K_F,\gamma)}
\ee
when $\gamma$ is close to $\gR$.

From Eqs.~(\ref{Kdeep}) and \eqref{eq:Delta} we find
\be
K_D-K_F = -\frac{\sqrt{\Delta}-\half\Delta}{2\mathrm{Fr}^2\cos^2\gamma} \approx -\frac{\sqrt{\Delta}}{2\mathrm{Fr}^2\cos^2\gamma}
\ee
since $\Delta$ tends to zero as $\gamma\to\gR$. 
From $\Omega'_+(K_F,\gamma)=0$ we obtain $\Omega''_+(K_F,\gamma) = 2(\mathrm{Fr}\cos\gamma)^3$, yielding
\be\label{Idelta}
I_+^D(\gamma)\approx - 2\pi\rmi \frac{ f(K_F)e^{K_FR\cos(\gamma-\phi_\beta)}}{\mathrm{Fr}\cos\gamma\sqrt{\Delta(\gamma)}}~~ \text{for }\gamma\to\gR.
\ee

We know that $\Delta(\gR)=0$. Provided the root of $\Delta$ at $\gamma=\gR$ is single, the singularity at $\gamma=\gR$ is of order $(\gamma-\gR)^{-1/2}$ and is integrable. The wave amplitude thus remains finite as long as $\Delta$ has a \emph{simple} root at $\gR$.

Infinite amplitudes are possible when $\Delta(\gR)=\Delta'(\gR)=0$, which is to say that $\Delta$ has a \emph{double} root at $\gR$. A little algebra reveals that this exactly matches the criterion %Eq.~
\eqref{psi} for a Doppler resonance to exist. This situation occurs for some value of $\gamma$ when $\tau=\tR$, yet in the presence of shear, other Doppler resonances can occur as well.

%%%%%%%%%%%%%%% S E C T I O N %%%%%%%%%%%%%%%%%%
\subsection{Diverging amplitude at resonance}

Let us consider the leading order contribution to the wave amplitude from (as an example) a wave of type $D$ when $\tau$ is very a very small but nonzero distance from $\taures$. When $\tau=\taures$, $\Delta$ has a double root at $\gamma=\tgR$. At the actual value $\tau$, $\Delta$ has a simple pole at $\gamma=\gR$, and we consider $\tau\to\taures$ and hence $\gR\to\tgR$.

Considering $\Delta$ as a function of $\gamma$ and $\tau$ we may Taylor expand,
\be
\Delta(\gamma,\tau) = \delta\tau \,\Delta_\tau(\tgR,\taures) + \half(\delta\gamma)^2\, \Delta''(\tgR,\taures) + ...
\ee
where $\delta\gamma=\gamma-\tgR$, $\delta\tau=\tau-\taures$, and we used that $\Delta(\tgR,\taures)=\Delta'(\tgR,\taures)=0$ and neglected sub-leading orders. A prime denotes differentiation with respect to $\gamma$, a subscript $\tau$ differentiation with respect to $\tau$. 
Inserting this into \eqref{zetaD} and \eqref{Idelta} gives the leading order contribution to the far-field wave as
\begin{align}
\frac{\zeta^D_{\mathrm{f.f.}}}{b}\propto & \int^{\gR} \frac{\rmd\gamma}{\sqrt{(\gamma-\gR)^2+2\frac{\Delta_\tau(\tgR,\taures)}{\Delta''(\tgR,\taures))}\delta\tau}}
\propto \ln(\delta\tau) + ...
\end{align}

We have shown that the Doppler resonance gives a logarithmically diverging wave amplitude, which is in agreement with the findings of \citet{dagan80,dagan82}.

%%%%%%%%%%%%%%% S E C T I O N %%%%%%%%%%%%%%%%%%
\subsubsection{Exception: finite resonance amplitude when $\tR=0$}

We find in the following that for certain velocity directions $\beta$ the smallest resonant value of $\tau$ can become identically zero. This only occurs for $\Frs\geq 1$, so if $\tau=\taures=0$ this must mean that $\omega_0=0$ but $V>0$ lest $\Frs=0$ as well.

Regarding the graphical dispersion relation solutions in Fig.~\ref{fig:dispRe-inf}, it is clear that if a wave solution of type $F$ exists when $\Omega_0=0$, this must imply $K_F=0$. Now notice that the function $f(\bK)$ in \eqref{f} tends to zero as $K\to 0$, which cancels the logarithmic divergence. This is in agreement with studies of ship waves (i.e., the case $\omega_0=0$) where no diverging amplitude is observed at the critical velocity \citep{ellingsen14a,li16}; quite the opposite, the amplitude of the transverse waves which become excluded tends to zero as velocity approaches critical. Note that this is not the case in two dimensions, where the waves made by a time-constant moving pressure distribution travelling at critical velocity gives rise to waves whose amplitude appears infinite until higher order terms are accounted for \citep{akylas84b}. Waves generated by a ship (or model of such) near critical velocity is a much studied problem, and upstream solitons are known to appear for transcritical velocities \citep[see, e.g.,][]{ertekin86,katsis87,lee89} particularly when the spanwise wave number is discretised by the presence of a channel of finite width.

%%%%%%%%%%%%%%% S E C T I O N %%%%%%%%%%%%%%%%%%
\subsection{Position of resonance frequencies}

In the following we determine the resonant values of $\tau$ for different values of $\beta$ and $\Frs$.

%%%%%%%%%%%%%%% S E C T I O N %%%%%%%%%%%%%%%%%%
\subsubsection{The minimal resonance $\tR$}\label{sec:taumin}

The smallest non-dimensional resonance frequency $\tR$ is the smallest value of $\tau$ so that $\tau=\Phi(\gamma)$ (see Eq.~\eqref{tauex}) has a solution, i.e., 
\be \label{eq:reson_f}
\tR=\min\limits_{\gamma}\left\lbrace \Phi(\gamma)\right\rbrace .
\ee
The notation denotes that the minimum value is found with respect to $ \gamma $ in the sector $ (-\pi/2, \pi/2) $. 

We shall see in the following that while $\tR$ is the smallest value at which infinite wave amplitudes may occur, it is not necessarily the only value. 
When shear current is not present, the well-known resonance found in the classical literature \citep{wehausen60} is obtained, i.e. $\tR=1/4$, which is then the only resonance. 

Based on Eq.~\eqref{eq:reson_f}, Fig.\ref{fig:resonance} presents the smallest resonance frequency $\tR$ for various $ \Frs $ and $ \beta $. The resonance frequency reaches its peak value at $ \beta=\pm \pi $ for given $ \Frs $, and its minimum for $ \beta=0 $. Moreover, the shear vorticity represented by $ \Frs $ tends to decrease the resonance frequency for $ -\pi/2<\beta<\pi/2 $. Notably, $\tR$ tends %rapidly 
to zero for increasing $\Frs$ when $ -\pi/2<\beta<\pi/2$, an observation which could well have implications for the 
heave and pitch %} 
of marine vessels in the presence of shear current, since resonance (often corresponding to sudden increase in wave loads) could occur at a much lower frequency than on still water or uniform current.

\begin{figure} 
	\graphicspath{{figures/}}
	\includegraphics[width=\textwidth]{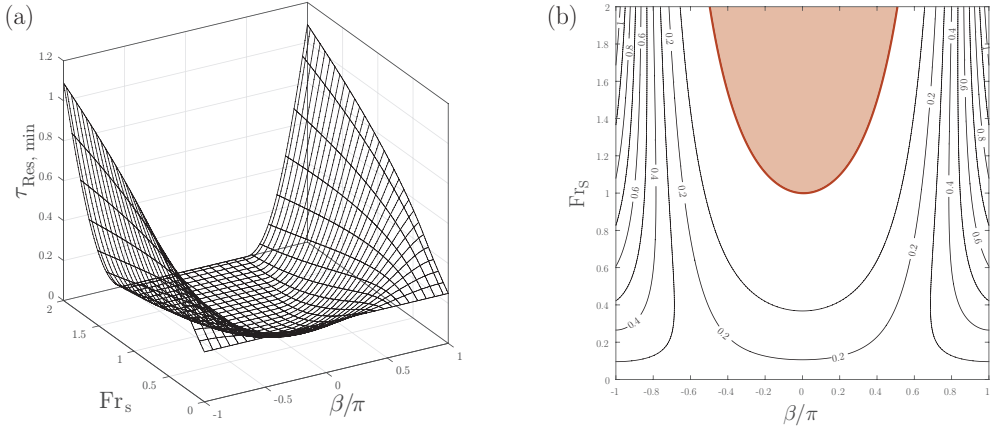}
	\caption{Smallest dimensionless resonance frequency $\tR$ as a function of $ \Frs $ and $ \beta $, (a) 3D plot and (b) contour plot. The shaded region satisfies Eq.~\eqref{zerotau}, and here $\tR=0$.}
	\label{fig:resonance}
\end{figure}
%%%%%%%%%%%%%%%%%%%%%%%%%%%%%%%%%%%%%%%%%%%%%%%%%%%%%%%%%%%%%%%%%%%%%%%%%%%%%%%

Resonance conditions in 2D 
were worked out by \citet{tyvand15} for 
shear assisted motion 
(corresponding to $\beta=\pi$ in 3D), in which case they obtain 
\be \label{eq_resf_2Dty}
\tR=(1+\Frs)^2/4.
\ee
The 2D geometry with the source travelling in the %opposite 
shear inhibited 
direction ($\beta=0$ in 3D) is not considered by \citet{tyvand15}.

We find that the 2D result \eqref{eq_resf_2Dty} is also a resonance of a 3D source moving in direction $\beta=\pi$, but it is not necessarily the smallest one. An explicit expression for $\tR$ in 3D is not available in general, but may be found from \eqref{eq:reson_f} 
when 
$\beta=0,\pm\pi$:
\bs \label{eq:res_0pi}
\begin{align}
\tR(\beta=\pm\pi)=&\left\lbrace \begin{array}{cl}
\frac14(1+\Frs)^2; &0\leqslant\Frs\leqslant\frac{1}{3}\\
\frac{4}{3}\sqrt{\frac13 \Frs}; &\Frs>\frac{1}{3}
\end{array}\right., \\
\tR(\beta=0)=&\left\lbrace \begin{array}{cl}
\frac14 (1-\Frs)^2; &0\leqslant\Frs\leqslant1\\
0; &\Frs>1
\end{array}\right. . 
\end{align}
\es
The smallest resonance frequency for $\beta=\pm\pi$ is smaller than the 2D result Eq.~\eqref{eq_resf_2Dty} when $\Frs>1/3$. The reason is that for $\beta=\pm\pi, \Frs>1/3$, the resonance condition $\Delta=0$ is first satisfied for a partial wave in directions 
\be
\gamma_\text{Res} = \pm \arctan\sqrt{3\Frs-1}
\ee
rather than  $\gamma=0$, as illustrated in Fig.~\ref{fig:excl}b. 

We plot the resonance frequencies as a function of $\Frs$ for different values of $\beta$ in Fig. \ref{fig:2dVS3d} for moderate values of $\Frs$. Some higher-than-minimal values of $\taures$ also appear in the figure, to be discussed in the next section. Note that the highest branch of resonant $\taures$ values at $\beta=\pi$ (marked with a circle in the figure) is the 2D result from \citet{tyvand15}, and that two different resonant values can occur quite close to each other in this case. 

%%%%%%%%%%%%%%%%%%%%%%%%%%%%%%%%%%%%%%%%%%%%%%%%%%
\begin{figure}
	\graphicspath{{figures/}}
	\centering{\includegraphics[width=5in]{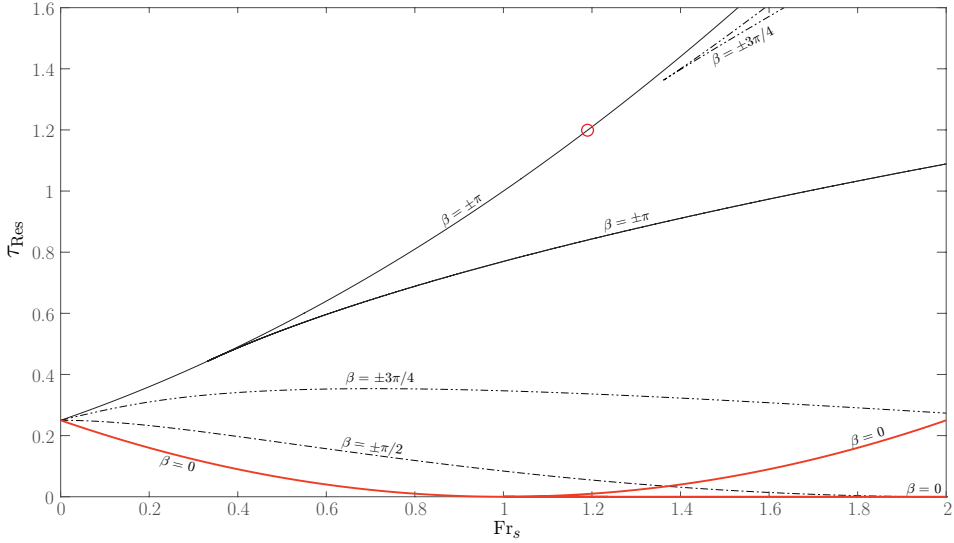}}
	\caption{Resonance frequencies $\taures$ in different directions of motion $\beta$ for moderate values of $\Frs$. The highest branch of resonant values for $\beta=\pm\pi$ (marked with a circle) corresponds to the 2D result of \citet{tyvand15}. }
	\label{fig:2dVS3d}
\end{figure}
%%%%%%%%%%%%%%%%%%%%%%%%%%%%%%%%%%%%%%%%%%%%%%%%%%

Except for the special case $\beta=\pm\pi$, the resonance frequency $\tR$ will always drop to zero for sufficiently high values of $\Frs$, because the numerator of Eq.~\eqref{eq:reson_f} will be zero for some $\gamma$. A bit of algebra shows that this is the case, and $\tR=0$, provided
\be\label{zerotau}
\Frs \geq 1/\cos^2(\beta/2).
\ee
This is exactly the criterion found for critical velocity to occur for ship waves (i.e., $\omega_0=0$) in deep water by \citet{ellingsen14a}. In Fig.~\ref{fig:resonance}b this region is shaded. As indicated by the $\beta=0$ graph in Fig.~ \ref{fig:2dVS3d}, when $\tR$ hits $0$ when $\Frs$ increases, it stays zero for all higher values, but a second branch of resonances also appears with increasing $\taures$ as a function of $\Frs$.

We finally note that for $\Frs\ll 1$, $\tR$ behaves like
\be
  \tR \sim \frac14\left[1-2 \Frs\cos\beta +\frac12\Frs^2(3\cos2\beta-1)+...\right].
\ee
Thus the minimal resonance changes linearly as a function of $\Frs$ for small shear, except when $\bV$ is orthogonal to the shear current ($\beta=\pm\pi/2$) when the behaviour is quadratic.

%%%%%%%%%%%%%%% S E C T I O N %%%%%%%%%%%%%%%%%%
\subsubsection{Additional resonances}\label{sec:respos}

\begin{figure}
	\graphicspath{{figures/}}
	\includegraphics[width=\textwidth]{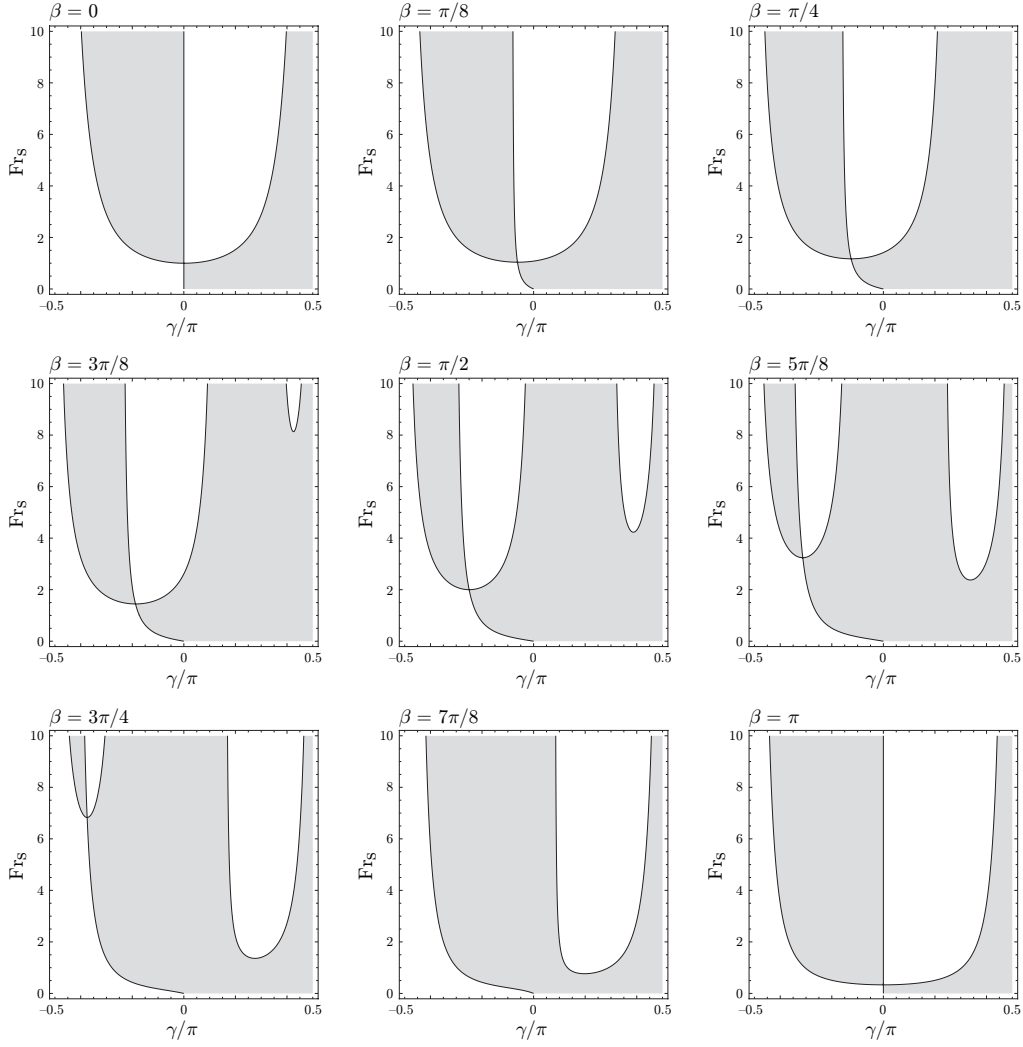}
	\caption{Solutions of $\Phi'(\gamma)=0$ from Eq.~\eqref{psi}. Shaded areas indicate $\Phi'(\gamma)>0$. To each solution of $\Phi'(\gamma)=0$ (at $\gamma=\gammares$, say) for a given $\Frs$ there exists a corresponding (not necessarily distinct) resonant $\tau$-value, $\taures=\Phi(\gammares)$, as plotted in Fig.~\ref{fig:res}.}
	\label{fig:psi}
\end{figure}

\begin{figure}
	\graphicspath{{figures/}}
	\includegraphics[width=\textwidth]{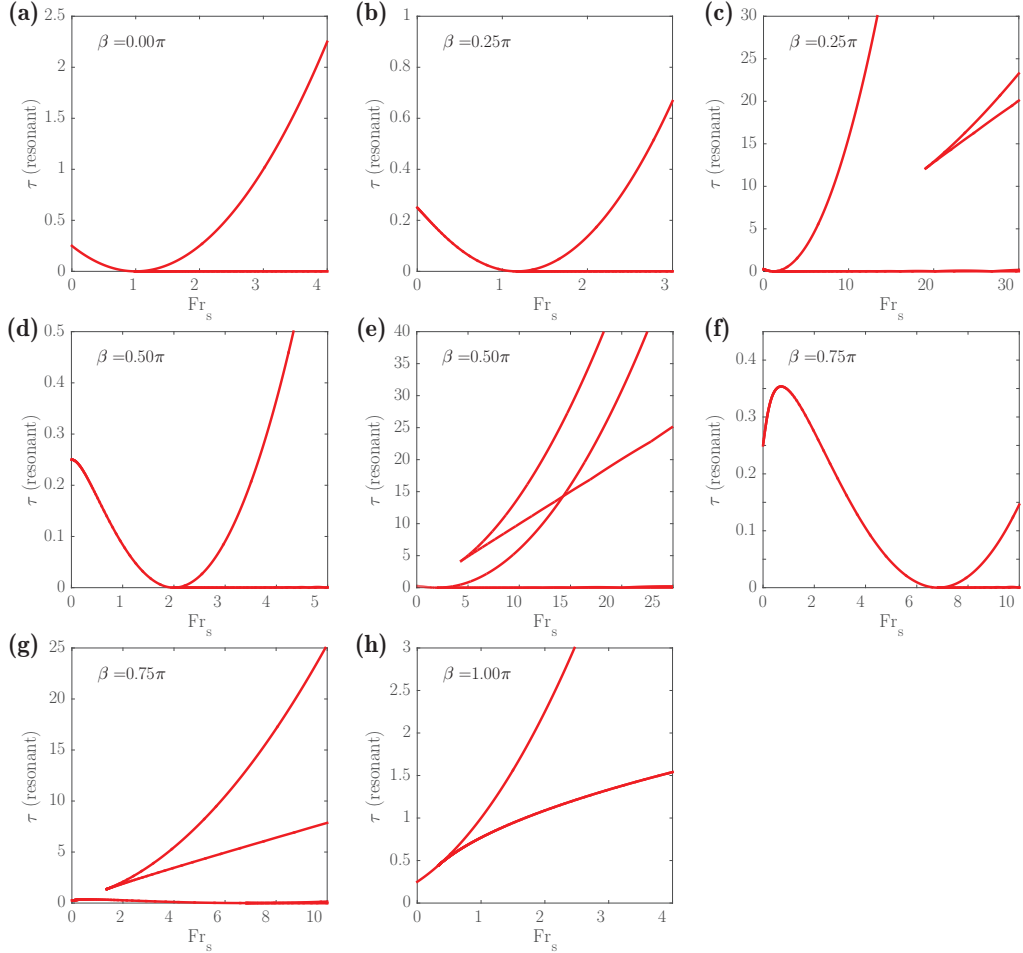}
	\caption{Doppler resonant frequencies $\taures$ as a function of $\Frs$ for different directions of motion, $\beta$. Where two panels show the same $\beta$ (Panels b and c, d and e, and f and g), the first is a zoom of behaviour for small and moderate $\Frs$ while the second shows the full picture appearing at large $\Frs$.}
	\label{fig:res}
\end{figure}

 As found in section \ref{sec:criterion}, the criterion for a Doppler resonance to exist is $\Phi'(\gamma)=0$ when $|\gamma|<\pi/2$. 
In other words, for some value of $\Frs$, there is a resonant (but not necessarily distinct) value $\tau=\Phi(\gamma)$ associated with each local maximum or minimum of $\Phi(\gamma)$;
if $\Phi(\gamma)$ has a local extremum at $\gamma=\gammares$, then $\tau=\Phi(\gammares)$ is a resonant value. In the absence of shear, resonance can only occur at $\gamma=0$. Fig.~\ref{fig:psi} shows that the introduction of a shear current results in a much richer resonance situation, where 
group velocity can vanish for different $\gamma$.

By mapping the zeros of $\Phi'(\gamma)$ onto the $\tau$ axis by requiring $\Delta(\gammares,\taures)=0$ we plot resonance frequencies $\taures$ as a function of $\Frs$ for different directions of motion, $\beta$, in Fig.~\ref{fig:res} (compare also with Fig.~\ref{fig:2dVS3d}). Panels a,b,d,f and h show the structure of resonances for low to moderate $\Frs$.
At larger values of $\Frs$, a more complex picture emerges, as shown clearly, e.g., in Fig.~\ref{fig:res}e. For $\Frs\gtrsim 4.2$, four different resonant values of $\tau$ can be identified (one of which is zero). 

\begin{figure}
	\graphicspath{{figures/}}
	\includegraphics[width=\textwidth]{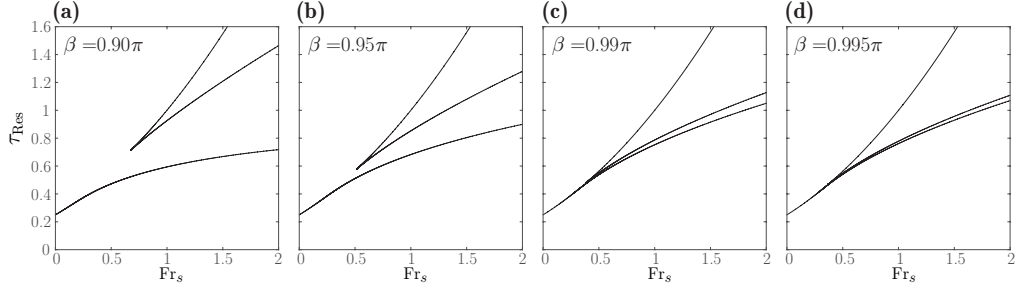}
	\caption{Doppler resonant frequencies $\taures$ as a function of $\Frs$ as $\beta$ approaches $\pi$ (or $ -\pi $). }
	\label{fig:restopi}
\end{figure}

Of particular interest is the situation for directions close to $\beta=\pm\pi$ as shown in Fig.~\ref{fig:restopi}. In Fig.~\ref{fig:restopi} we plot the structure of resonances for directions close to this maximally shear inhibited direction. As, $\beta\to\pi$, the example shown, a pair of higher resonances begin to appear at ever lower $\Frs$, the smaller of which eventually merging with the smallest resonance with the branch point reaching $\Frs=1/3$. 

%%%%%%%%%%%%%%%%%%%%%%%%%%%%%%%%%%%%%%%%%%%%%%%%%%%%%%%%%%%%%%%%%%%%
%%%%%%%%%%%%%%%%%%%%%%%%% S E C T I O N %%%%%%%%%%%%%%%%%%%%%%%%%%%%
%%%%%%%%%%%%%%%%%%%%%%%%%%%%%%%%%%%%%%%%%%%%%%%%%%%%%%%%%%%%%%%%%%%%
\section{Concluding remarks}

We have studied the waves on the free surface atop a shear current which, when undisturbed, has uniform vorticity. The wave-making perturbation, modelled as a surface pressure distribution, is at one time oscillating in strength and moving relative to the free surface with constant velocity making an arbitrary angle $\beta$ with the sub-surface shear current. 
In the absence of vorticity the problem is a classical one, with applications in the study of ship motion in regular waves.

We provide a detailed analysis of the dispersion relation which must be fulfilled for 
far-field waves. Both finite and infinite water depth are considered. Graphical solution of the dispersion relation reveals a considerably more complex picture than was the case when no shear current is present. As has long been known for the still water case, for values of $\tau=|\bV|\omega_0/g$ less than a critical value [$\bV$ is disturbance velocity, $\omega$ is disturbance frequency, $g$ is the acceleration of gravity], we find for finite water depth always four waves, three of which behind the disturbance and one travelling ahead. Above the smallest resonant value, $\tR$, two of these waves, including the forward propagating one, vanish from at least one sector of propagation directions. 

The situation is far richer than in the absence of shear, however. Firstly, several resonance frequencies (as many as $4$) can occur for some sets of parameters, and, correspondingly, several sectors of wave propagation directions can exist wherein no far-field wave solutions exist. Secondly, in deep waters the phenomenon of ``cut-off'' reported by \cite{tyvand14} occurs if $\omega_0<S$ [$S$ is the vorticity of the undisturbed flow], in which case one of the wave solutions effectively disappears within a sector of shear assisted propagation directions.

The structure of resonant values of $\tau$ is analysed thereafter, revealing a complex picture where up to $4$ different resonant values of $\tau$ can exist for any combination of parameters $\beta$ and the ``shear-Froude number'' $\Frs=VS/g$. The situation is particularly notable when the disturbance motion is close to maximally shear inhibited ($\beta=0$) or shear assisted ($\beta=\pm\pi$). In the former situation, $\tR$ decreases rapidly towards zero for increasing values of $\Frs$ and splits into separate branches for $\Frs>1$, one of which is $\tR=0$. In the latter situation, when $\beta$ is close to, but not exactly, $\pm\pi$, three resonant values exist for $\Frs$ greater than some critical value $1/3<\Frs\lesssim 1$. For directions of motion close to orthogonal with the shear flow, the same richness of resonances exists, but requires values of $\Frs$ well in excess of $1$, which may be difficult to achieve in practice.

For all directions of motion that are significantly assisted or inhibited by the shear, the resonant value of $\tau$ changes rapidly, as $\tR\sim \frac14(1-2\Frs\cos\beta+...)$ for $\Frs\ll 1$. Thus the presence of a shear current will change the resonant value significantly even for $\Frs\sim \mathcal{O}(10^{-1})$.

\bibliographystyle{jfm}
\bibliography{references}

\end{document}